\documentstyle[aps,prl,twocolumn]{revtex}
\include{psfig}
\begin{document}
\draft

\title {Instantaneous Normal Mode Analysis of Supercooled Water}

\author{Emilia La Nave$^1$, Antonio Scala$^1$,  Francis~W. 
Starr$^{1}$\thanks{Current
Address: Polymers Division and Center for Theoretical and Computational
Materials Science, National Institute of Standards and Technology,
Gaithersburg, Maryland 20899}, \\
Francesco Sciortino$^{2}$, and H. Eugene Stanley$^{1}$}

\address{$^1$Center for Polymer Studies, Center for Computational
Science, and Department of Physics, Boston University, Boston, MA
02215 USA}

\address{$^2$Dipartmento di Fisica e Istituto Nazionale per la Fisica
della Materia, Universit\'{a} di Roma ``La Sapienza'', Piazzale Aldo
Moro 2, I-00185, Roma, Italy}

\date{\today}
%\date{Revised: \today; ~~~Printed: \today; ~~~lssss.tex}

\maketitle
\begin{abstract}
We use the instantaneous normal mode approach to provide a description
of the local curvature of the potential energy surface of a
model for water.  We focus on the region of the phase diagram in which
the dynamics may be described by the mode-coupling theory.  We find,
surprisingly, 
that the diffusion constant depends mainly on the fraction of directions
in configuration space connecting different local minima, supporting the
conjecture that the dynamics are controlled by the geometric properties
of configuration space.  Furthermore, we find an unexpected relation
between the number of basins accessed in equilibrium and the
connectivity between them. 
\end{abstract}
\bigskip

%\begin{multicols}{2}

The study of the dynamics in supercooled liquids has received great
interest in recent years\cite{debenedetti} 
due to possibilities opened by novel
experimental techniques~\cite{exptec}, detailed theoretical
predictions~\cite{mct}, and by the possibility of following the
microscopic dynamics via computer simulation~\cite{walter-review}.  One
important theoretical approach developed recently is the ideal mode
coupling theory(MCT)~\cite{mct}, which quantitatively predicts the time
evolution of correlation functions and the 
dependence on temperature $T$ 
of characteristic correlation times.  Unfortunately, the temperature
region in which MCT is able to make predictions for the long time
dynamics is limited to weakly supercooled states~\cite{note-weakly}.  In
parallel with the development of MCT, theoretical
work~\cite{parisi,speedy,swarz,wallace} has called attention to
thermodynamic approaches to the glass transition, and to the role of
configurational entropy in the slowing down of dynamics. These theories,
which build on ideas put forward some time
ago~\cite{stillinger,jones,goldstein}, stress the relevance of the
topology of the potential energy surface (PES) 
explored in supercooled states.
Detailed studies of the PES may provide insights on the
slow dynamics of liquids, and new ideas for extending the present
theories to the deep supercooling regime.

The instantaneous normal mode (INM) approach uses the
eigenvectors (normal modes) and eigenvalues of the Hessian matrix (the
second derivative of the potential energy) to provide an
``instantaneous'' representation of the PES
in the neighborhood of an equilibrium phase space point.  Examples of
INM analysis have appeared for a few representative
liquids~\cite{keys2,madden,ohmina,scitar}. In this Letter 
 we present INM calculations
for a rigid water molecule model, the SPC/E potential~\cite{berensen},
for several temperatures and densities in supercooled states.  The
reasons for selecting such a system are twofold: (i) The SPC/E model has
been studied in detail, and it has been shown that its dynamics follows
closely the predictions of MCT~\cite{roma,francis}.  Furthermore, the
($\rho$, $T$) dependence of the configurational entropy has been
calculated, and shown to correlate with the dynamical
behavior~\cite{antonio}. (ii) The SPC/E model has a maximum of
the diffusion constant $D$ under pressure along isotherms, as observed
experimentally~\cite{prielmeier}.  
The maxima in $D$ can be used as sensitive probes for
testing proposed relations between the change in topology of
configuration space and the change in $D$.

We analyze recent simulations~\cite{francis} for six densities between
$0.95$ and $1.3$~g/cm$^3$ and five temperatures for each density.  For
each state point, we extract 100 equally-spaced
configurations~\cite{CORREL-note}, from which we calculate and
diagonalize the Hessian using the center of mass and the three principal
inertia momenta vectors as molecular
coordinates~\cite{note-coordinates}. For each configuration, we classify
the imaginary modes into two categories: (i) shoulder modes and (ii)
double-well modes, according to the shape of the PES along the
corresponding eigenvector~\cite{laird}.  The shoulder modes are modes
for which the negative curvature appears as a result of a local
anharmonicity; the double-well modes include only the directions in
configuration space which connect two adjacent minima.  The
classification of the modes is done by studying the shape of the PES 
along the eigenvector corresponding to the mode, i.e. beyond the
point in configuration space where such modes have been calculated.  In
this respect, the potential energy along the eigenvector direction may
be different from the actual curvilinear direction described by the time
evolution of the eigenvector under consideration.  Furthermore, some
modes classified as double-well may be related to intra-basin
motions~\cite{gezelter} and therefore may not be relevant to describe
diffusivity. In the present study we show --- using the $D$ maxima line
as a guide --- that these spurious features are negligible.

\begin{figure}[htbp]
\begin{center}
\mbox{\psfig{figure=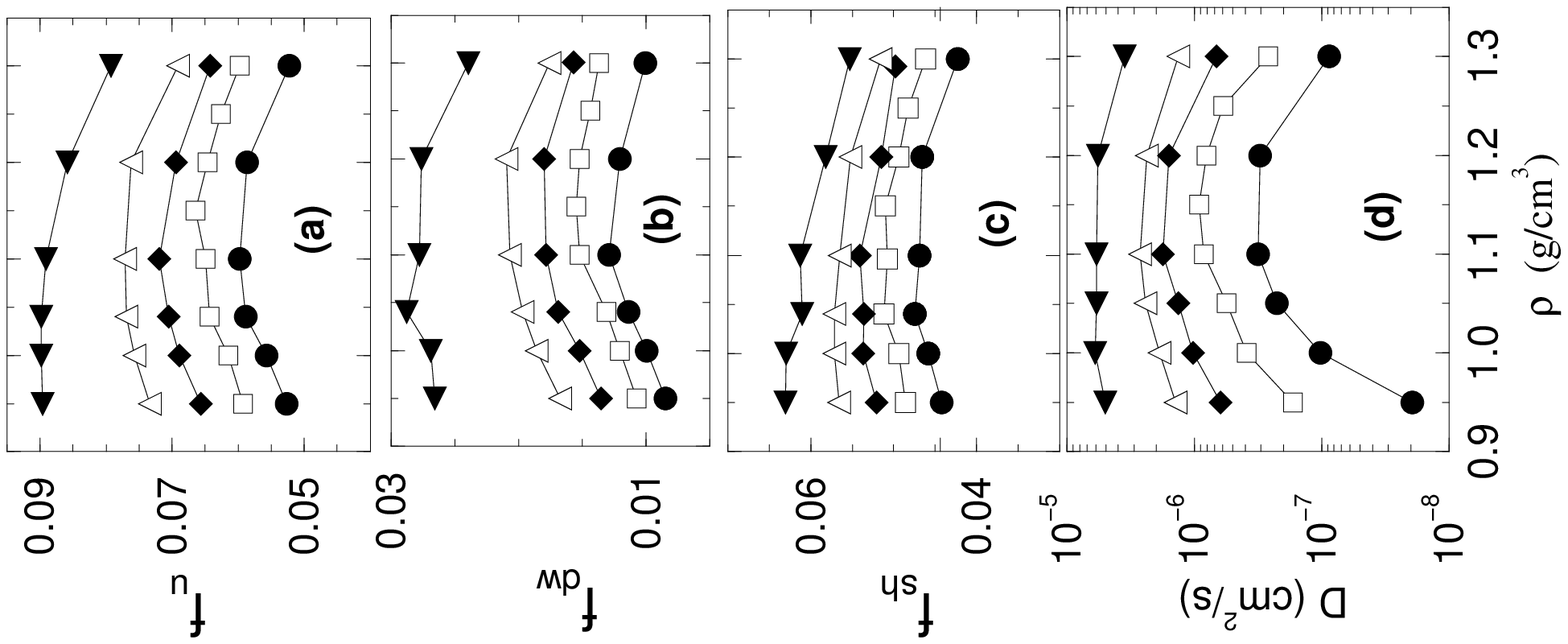,width=9cm,angle=-90}}
\end{center}
\caption{Density dependence along the studied isotherms of
(a) $f_u$, (b) $f_{dw}$, (c) $f_{sh}$. 
Temperatures are $T=210 K$ (circles), $T=220 K$ (squares), $T=230 K$ 
(diamonds) and  $T=240 K$ (triangles). (d) shows the $\rho$ dependence of
$D$ for the same isotherms~\protect\cite{francis}.}
\label{fig:data}
\end{figure}

\begin{figure}[htbp]
\begin{center}
\mbox{\psfig{figure=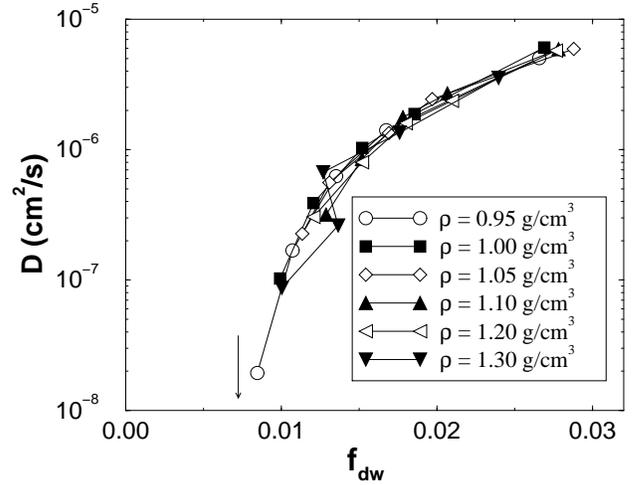,width=8cm,angle=-90}}
\end{center}
\caption{Diffusion constant $D$  versus $f_{dw}$ for different isochores.
The arrow indicates the value $f_{dw} \approx 0.007$, to 
highlight the presence --- even when $D$ approaches zero ---
of a small number of spurious double-well modes, related to intrabasin 
motion\protect\cite{gezelter,note-insidedw}.}
\label{fig:dvsfdw}
\end{figure}

Fig.~\ref{fig:data} shows the density dependence of the fractions  
of
imaginary (unstable) modes $f_u$, double-well modes $f_{dw}$, and of
shoulder modes $f_{sh}$, where $f_u=f_{dw}+f_{sh}$; we also show $D$ for
the same state points.  The close correlation of $f_{dw}$ and $D$ is
striking.  At $\rho\approx 1.15$~g/cm$^3$, $D$ has a maximum; at this
$\rho$, the increase of diffusion constant caused by the the progressive
disruption of the hydrogen bond network on increasing the density is
balanced by slowing down of the dynamics due to increased packing.  At
the same density, $f_u$ and $f_{dw}$ also show maxima, supporting the
view that these quantities are a good indicator of the molecular
mobility.  There is also a weak maximum in $f_{sh}$, but at a $\rho <
1.15$~g/cm$^3$\cite{note-dist}.  
The presence of maxima in both $D$ and $f_{dw}$ at the same density
suggests that, for the SPC/E potential, $f_{dw}$ is directly related to
$D$.  The relation between $D$ and $f_{dw}$ is shown in
Fig.~\ref{fig:dvsfdw} for all the studied isochores.  
We note that $D$ in this system is a monotonic
function of $f_{dw}$ and that all points fall on the same master
curve\cite{note-keyes}, notwithstanding the large range of $D$ values 
analysed. Thus, surprisingly, in SPC/E water, 
the knowledge of the fraction of directions
in configuration space leading to a different basin, $f_{dw}$, is
sufficient to determine the value of $D$.\cite{material}  
We also note from
Fig.~\ref{fig:dvsfdw} that $D$ vanishes, but at a small non-zero value
of $f_{dw} \approx 0.007 $ (vertical arrow)~\cite{estimate}, suggesting
that indeed a small number of spurious double-well modes, related to
intrabasin motion~\cite{gezelter,note-insidedw}, are still included in
our classification.  

The reduction of mobility on cooling in the studied $(\rho,T)$ range --
where MCT provides a description of the dynamics -- appears related
to the geometrical properties of the PES, i.e. the system mobility is
reduced because the number of directions connecting different local
minima (needed to explore {\it freely} the configuration space) is
decreasing~\cite{campbell,p-spin}.  Hence, the observed reduced mobility is
mainly related to the geometry of the PES, i.e. it is ``entropic'' in origin.

\begin{figure}[htbp]
\begin{center}
\mbox{\psfig{figure=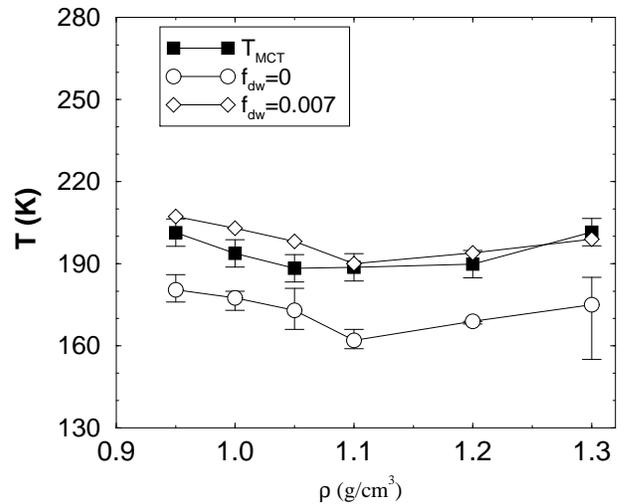,width=8cm,angle=-90}}
\end{center}
\caption{Comparison of the mode coupling critical temperature $T_{MCT}$
\protect\cite{francis}, of the $f_{dw}=0$ locus and of 
the $f_{dw}=0.007$ locus.}
\label{ZZ}
\end{figure}

MCT seems to be able to describe the entropic slowing down of the
dynamics associated with the vanishing of $f_{dw}$.  This is consistent
with the general consensus that the missing decay channels for the
correlations responsible for the failure of MCT at very low $T$ are
activated processes.  To make closer
contact with MCT, we compare in Fig.~\ref{ZZ} the density dependence of
the MCT critical temperature $T_{MCT}$~\cite{francis} with the $T$ at
which the fraction of double-well modes goes to zero, as well as with
the $T$ at which $f_{dw}$ reaches the estimated asymptotic value of
$0.007$~\cite{note-insidedw}, to take into account the small over counting
intrinsic in the mode classification.  We observe that the $f_{dw}=0$
locus tracks closely the $T_{MCT}$ line, but that the $f_{dw}=0.007$
locus nearly coincides with the $T_{MCT}$ line.

\begin{figure}[htbp]
\begin{center}
\mbox{\psfig{figure=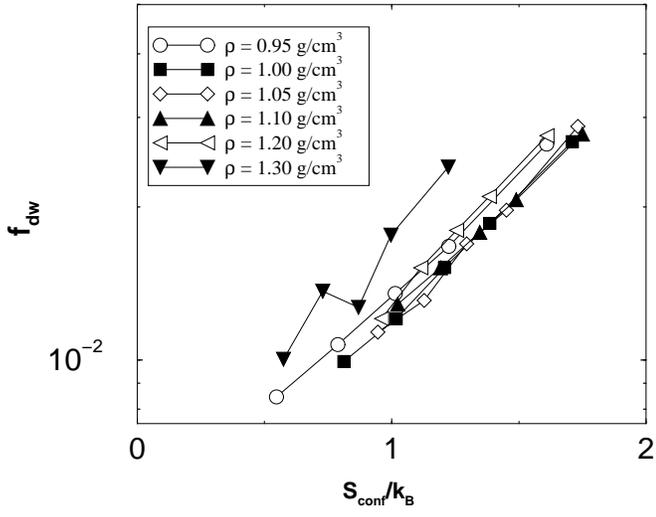,width=8cm,angle=-90}}
\end{center}
\caption{Linear-log plot of $S_{\mbox{\scriptsize conf}}$ 
(from Ref.\protect\cite{antonio}) vs
$f_{dw}$.}
\label{fig:fig4}
\end{figure}

The results presented here suggest that the liquid dynamics in the MCT
region of SPC/E water is controlled by the average connectivity in
configuration space.  On the other hand, for the same model, the
($\rho$, $T$) dependence of the configurational entropy
$S_{\mbox{\scriptsize conf}}$ -- which, in the 
Stillinger-Weber~\cite{stillinger} formalism,  
can be defined as the logarithm of
the number of different basins $\Omega$ in configuration space -- has
also been calculated and shown to correlate with
$D$\cite{antonio}. Since no {\it a priori} relation is expected between
the connectivity of the local minima and their number, we consider
particularly interesting the observation of such a relation.
For this reason, we show in
Fig.~\ref{fig:fig4} a ``parametric plot'' (in the parameter $D$) of
$f_{dw}$ versus $S_{\mbox{\scriptsize conf}}$ for all studied densities.
 Fig.~\ref{fig:fig4} shows a linear relation between $\ln
(f_{dw})$ and $S_{\mbox{\scriptsize conf}}$, i.e. between $f_{dw}$ and the number of basins
(since $\Omega=\exp (S_{\mbox{\scriptsize conf}}/k_B)$) and highlight
the existence of an unexpected relation between the number of basins
accessed in equilibrium and the connectivity between them.  This novel
feature of the PES contributes to a better understanding of the water
dynamics, and may contribute to the understanding of the dynamics of
glass-forming liquids\cite{novel}.  
%didn't like this?  ok... :(
%Furthermore, the relations may indicate the
%utility of studying the PES as a network, such as might be described by
%recent studies of small-world networks~\cite{small-world}.

We thank T.~Keyes for useful discussions, and the NSF for support; FS
acknowledges partial support from MURST (PRIN 98).

%\end{multicols}

%\end{multicols}


\begin{references}
\bibitem{note-weakly} 
We call ``weakly supercooled states'' the temperature range where the 
diffusivity has decreased 3 or 4 orders of magnitude compared to its 
normal liquid value ( $\sim10^{-5}cm^2/s$ ).
\bibitem{note-coordinates}
Ref.~\protect\cite{furkers} has called attention to the fact that in molecular
systems, the frequency distribution and the eingvectors are
dependent on the coordinate system chosen.  We refer the interested
reader to ref.\protect\cite{furkers} for more detailed discussions on this
point.
\bibitem{note-dist}
Thus far, we cannot distinguish if the shift in the
maximum of $f_{sh}$ is connected to a change of the anharmonicity as a
function of density, or to spurious leakage of double-well modes into
shoulder modes in the classification procedure.
\bibitem{note-keyes}
We note that for $D \gtrsim 0.3\times 10^{-6}$~cm$^2$/s, the
relation between $D$ and $f_{dw}$ is approximately linear, in agreement
with the finding of Keyes and coworkers for different
liquids~\protect\cite{keys2,keyes3}.
\bibitem{material} The observed relation $D(f_{dw})$ is material
dependent. Studies of the INM spectrum in different materials 
have shown that  
$D$ may correlate with $f_{u}$ or with the product $\omega_u f_u$, 
where $w_u$ is defined as the average frequency of the 
$f_u$ distribution\protect\cite{keys2,madden}.
\bibitem{estimate} The approximate value 0.007 has been
obtained by fitting $D$ vs $f_{dw}$ with a cubic polynomia
and evaluating the value at which $D=0.0$.
\bibitem{note-insidedw}
Ref.\protect\cite{gezelter} has shown that some of the
eigenvector of the double well type may indeed point, in both
directions, to the same local minimum; 
an example is the Mexican hat surface.
\end{references}
\end{document}